\documentclass[prl,aps,twocolumn,floats,superscriptaddress]{revtex4}
\usepackage{amsmath}
\usepackage{graphicx}
\usepackage{epsfig}
\newlength{\upit}\upit=0.1truein

\newcommand{\ltappr}{{{\lower4pt\hbox{$<$} } \atop \widetilde{ \ \ \ }}}
\newlength{\bxwidth}\bxwidth=1.5 truein

\begin{document}
\newcommand{\dg}{^{\dagger }}
\newcommand{\si}{\sigma}
\newcommand{\rarrow}{\rightarrow}
\def\fig#1#2{\includegraphics[height=#1]{#2}}
\def\figx#1#2{\includegraphics[width=#1]{#2}}
\newlength{\figwidth}
\newlength{\shift}
\shift=0.4cm
\newcommand{\fg}[3]
{
\begin{figure}[ht]
\vspace*{-0cm}
\[
\includegraphics[width=\figwidth]{#1}
\]
\vspace*{\shift}
\caption{\label{#2}
\small
#3
}
\end{figure}}
%\psdraft

\newcommand \bea {\begin{eqnarray} }
\newcommand \eea {\end{eqnarray}}

\title{Non Fermi Liquid behavior in the under-screened Kondo model}
\author{P. Coleman}

\affiliation{$^{2}$ Center for Materials Theory,
Rutgers University, Piscataway, NJ 08855, U.S.A. }

\author{C. P{\'e}pin}

\affiliation{SPhT, L'Orme des Merisiers, CEA-Saclay, 91191
Gif-sur-Yvette France.}

\begin{abstract}
Using the Schwinger boson spin representation, we reveal
a new aspect to the physics of a partially screened magnetic moment in
a metal, as described by the spin $S$ Kondo model.
We show that the  residual ferromagnetic interaction between a partially
screened spin 
and  the electron sea destabilizes 
the Landau Fermi liquid, forming a singular Fermi liquid 
with a $1/ (T \ln ^{4 }
(T_{K}/T))$ divergence in the low temperature specific heat
coefficient $C_{V}/T$. A magnetic field $B$ tunes this 
system back into Landau Fermi liquid with a Fermi temperature
proportional to $B \ln^2 (T_K/B)$. We discuss a possible link
with field-tuned quantum criticality in heavy electron materials. 
\end{abstract}
%\eject
%
\maketitle
%
%\vfill\eject 
Heavy electron materials are the focus of  renewed attention,
because of the opportunity\cite{stewartrmp,varma,qcp} they present to understand the
physics of matter near a quantum critical point. 
One of the unexplained properties of these materials, is that 
the characteristic temperature 
scale of heavy electron Fermi liquid is driven to zero at 
quantum critical point\cite{hvl,grosche,devisser,knebel,gegenwart}.  When either the paramagnet or
antiferromagnetic heavy electron phase is warmed above this temperature
scale, it enters a ``non-Fermi liquid'' phase. These results 
suggest that insight into the non-Fermi liquid behavior of heavy electron systems
might be obtained by studying the break-up of the antiferromagnetic
state.

Traditionally, ordered moment antiferromagnetism is described using a
bosonic representation of the ordered moments.  In this paper we
examine the under-screened Kondo impurity model (UKM) and we
demonstrate that the essential physics of the under-screened Kondo
effect is captured by a Schwinger boson representation of the local
moments.  In the course of our study we obtained an unexpected new
insight.  The UKM describes the screening of a local moment from spin
$S$ to spin $S^{*}= S -\frac{1}{2}$\cite{mattis}. At low temperatures,
this residual moment ultimately decouples from the surrounding Fermi
sea.  The UKM has been studied using the strong coupling
expansion\cite{noz}, 
numerical renormalization
group\cite{Cragg} and 
diagonalized  using the Bethe
Ansatz\cite{tsvelik,natan}, but 
the possibility of a break-down of 
Landau Fermi liquid behavior was not addressed.
In this paper, we
show that a field-polarized under-screened moment forms a Fermi liquid 
with a  the Fermi temperature that is proportional
to the magnetic field, going to zero when the field is removed. At zero field,
the residual coupling of the electron fluid 
to the  degenerate states of the under-screened moment violates the
phase space restrictions  required for formation of a Landau Fermi
liquid, leading to 
a 
strongly divergent specific heat coefficient (Fig. \ref{fig1}.)  
\begin{equation}\label{}
\frac{C_{V}}{T} \sim \frac{1}{ T \ln ^{4} (T_{K}/T)}.
\end{equation}
The
appearance of a field-tuned Fermi temperature is
strikingly reminiscent of the heavy electron quantum critical
points.\cite{grosche,gegenwart,schroeder,gegenwartnew} 

The UKM is model is written
\begin{eqnarray}\label{start}
H&=& \sum_{k\alpha } \epsilon _{k}c\dg _{k\sigma }c_{k\sigma
}+ J \vec S\cdot \psi \dg _{\alpha} \vec \sigma_{\alpha \beta} 
\psi _{\beta}
.
\end{eqnarray}
where $S$ denotes a spin $S>\frac{1}{2}$, $c\dg _{k \alpha }$ creates
a conduction electron with wave vector $k$, spin component $\alpha$,
$\psi\dg _{\alpha }= \sum_{k}c\dg_{k\alpha }$ creates a conduction
electron at the impurity site.  
\shift=-1cm
\figwidth=\columnwidth \fg{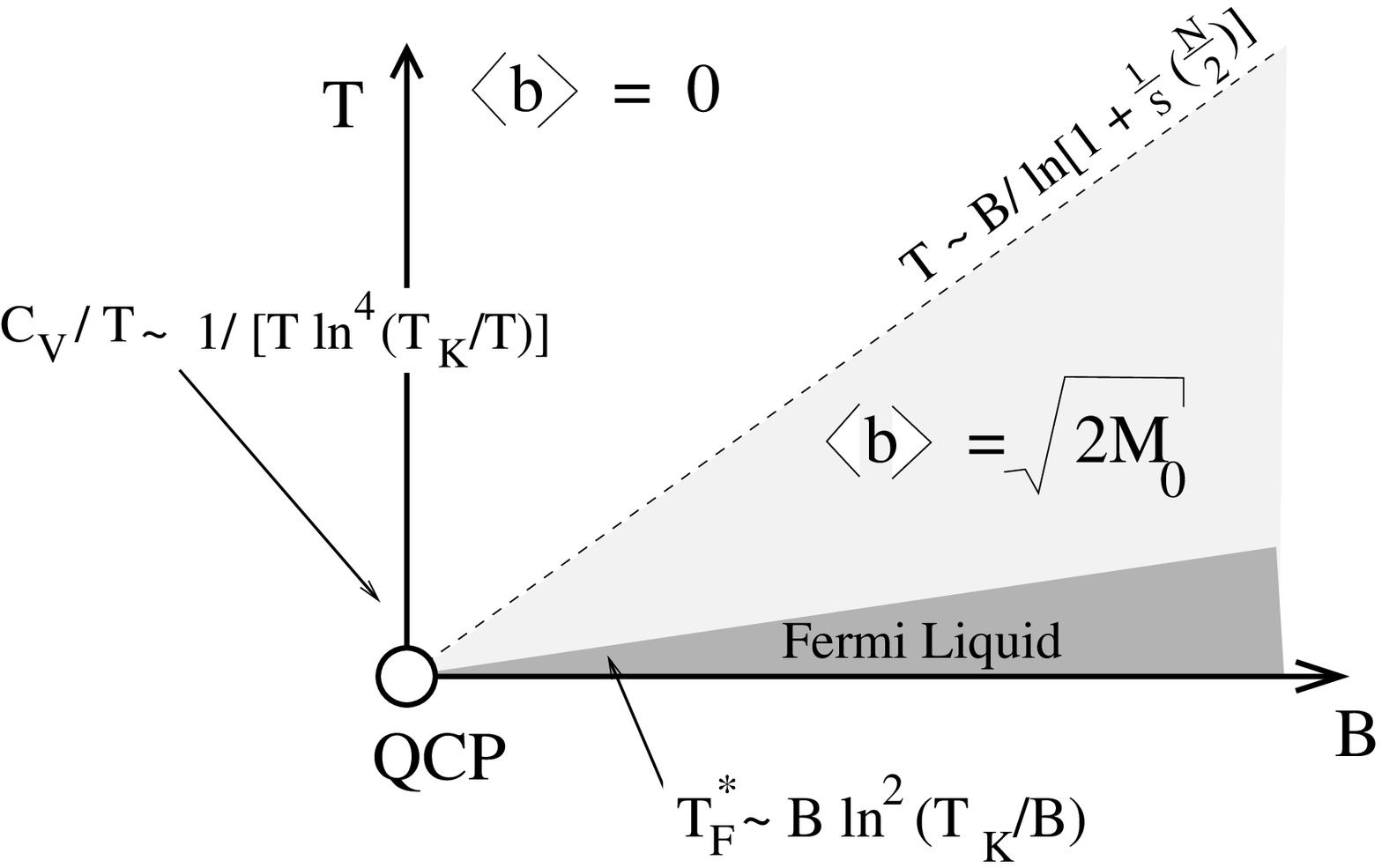}{fig1}{Schematic phase diagram
of the large $N$ limit of the under-screened Kondo model.
} 
\shift=-0.4cm

%Traditionally, the magnetic moment in the Kondo model
%is represented by a  bilinear combination of `` Abrikosov
%pseudo-fermions''. \cite{abrik} An expansion in $1/N$ around
%$N= \infty $ in this scheme 
%provides a good mean-field description of the paramagnetic
%Fermi liquid ground-state.
\noindent 
We begin by reformulating the UKM
as an SU(N) invariant Coqblin Schrieffer
model, which enables us to carry out a large $N$ expansion of the
physics. We write
\begin{eqnarray}\label{H1}
H&=& \sum_{k\alpha } \epsilon _{k}c\dg _{k\alpha }c_{k\alpha
}+\frac{J}{N}\sum_{\alpha \beta }(\psi \dg _{\beta }b_{\beta
})
(b\dg _{\alpha } \psi _{\alpha }) 
 - MB,
\end{eqnarray}
where the spin indices run over $N$ independent values $\alpha ,\beta
\in (1,N)$, and the constraint $n_b=2S$ is imposed to represent spin
$S$.  $M= g_{N}[b\dg _{\uparrow }b_{\uparrow}- \tilde{S}] $ is the
local moment magnetization, where we denote the first spin component
by $\sigma =1\equiv \uparrow$, and $\tilde{S}=\frac{2S}{N}$. The
pre-factor $g_{N}=\frac{N}{2(N-1)}$ is chosen so that at maximum
polarization, when $n_{\uparrow}=2S$, $M=S$. A multi-channel
formulation of the above model, has previously been treated within an
integral equation formalism\cite{parcollet97a}.

Next, we cast the partition function as a path integral and  factorize the interaction
\begin{equation}\label{}
H_{I}\rightarrow 
\left[
\bar \phi\  b\dg _{\sigma }\psi _{\sigma } +\psi \dg
_{\sigma }b_{\sigma }\phi 
\right]
-\frac{N}{J}\bar \phi \phi.
\end{equation}
We shall show how the physics of the under-screened Kondo model
is obtained by examining 
the Gaussian fluctuations of the field $\phi $
about the mean-field theory obtained by taking 
$N\rightarrow \infty $ at fixed $\tilde{S}$. 
This mean-field theory describes a free moment, with Free energy 
\begin{eqnarray}\label{}
F_{LM}&=& \sum _{\sigma =1}^{N} T\ln [1-e^{-\beta
(\lambda-\delta _{\sigma \uparrow}\tilde{B})}]- 2
(\lambda-\frac{\tilde{B}}{N})S.
\end{eqnarray}
where $\tilde{B}=g_NB$. The mean-field constraint $\langle
n_{b}\rangle =2S$ becomes
\begin{equation}
\langle n_{b}\rangle = n[\lambda-\tilde{B}] + (N-1)n[\lambda ]= 2S,
\end{equation}
where $n[x]= [e^{\beta x}-1]^{-1}$ is the Bose-Einstein distribution
function.  There are then two types of mean-field solution: 
\begin{enumerate}

\item ``Paramagnet''
where $\langle b_{1}\rangle =0$ and
$n(\lambda)= \tilde{S}$. 

\item ``Polarized'' moment where 
$\langle b_{1}\rangle = \sqrt{2M}$ condenses to 
produce magnetization $M$, 
\end{enumerate}
The second  phase develops at temperatures below 
$T=T_c = B/\zeta$, $\zeta=\ln[1 + \frac{1}{\tilde{S}}]$ 
(Fig. \ref{fig1}). The mean-field value of $\lambda$ in these two phases
is given by $\lambda=\lambda_{0}=\max (T \zeta, \tilde{B})$. 

To examine the fluctuations around the mean-field theory, we
integrate out the electrons and bosons, and write the effective
action so obtained in terms of the
Fourier coefficients $\phi_n = \beta ^{-1/2}
\int_0^{\beta}d\tau \phi(\tau)e^{i\omega_n \tau}$. To quadratic order,
order, the effective action is given by
\begin{eqnarray}\label{basic}
S[\bar \phi , \phi ] = - \sum_{\omega_{n}} \bar \phi_{n}
{\cal J}^{-1}_{n}
\phi_{n}.
\end{eqnarray}
Th propagator  ${\cal  J}_{n}$ for the $\phi $ field is determined
by the Feynman diagrams shown in Fig. \ref{fig2}(a).
\shift=-1cm
\figwidth=\columnwidth \fg{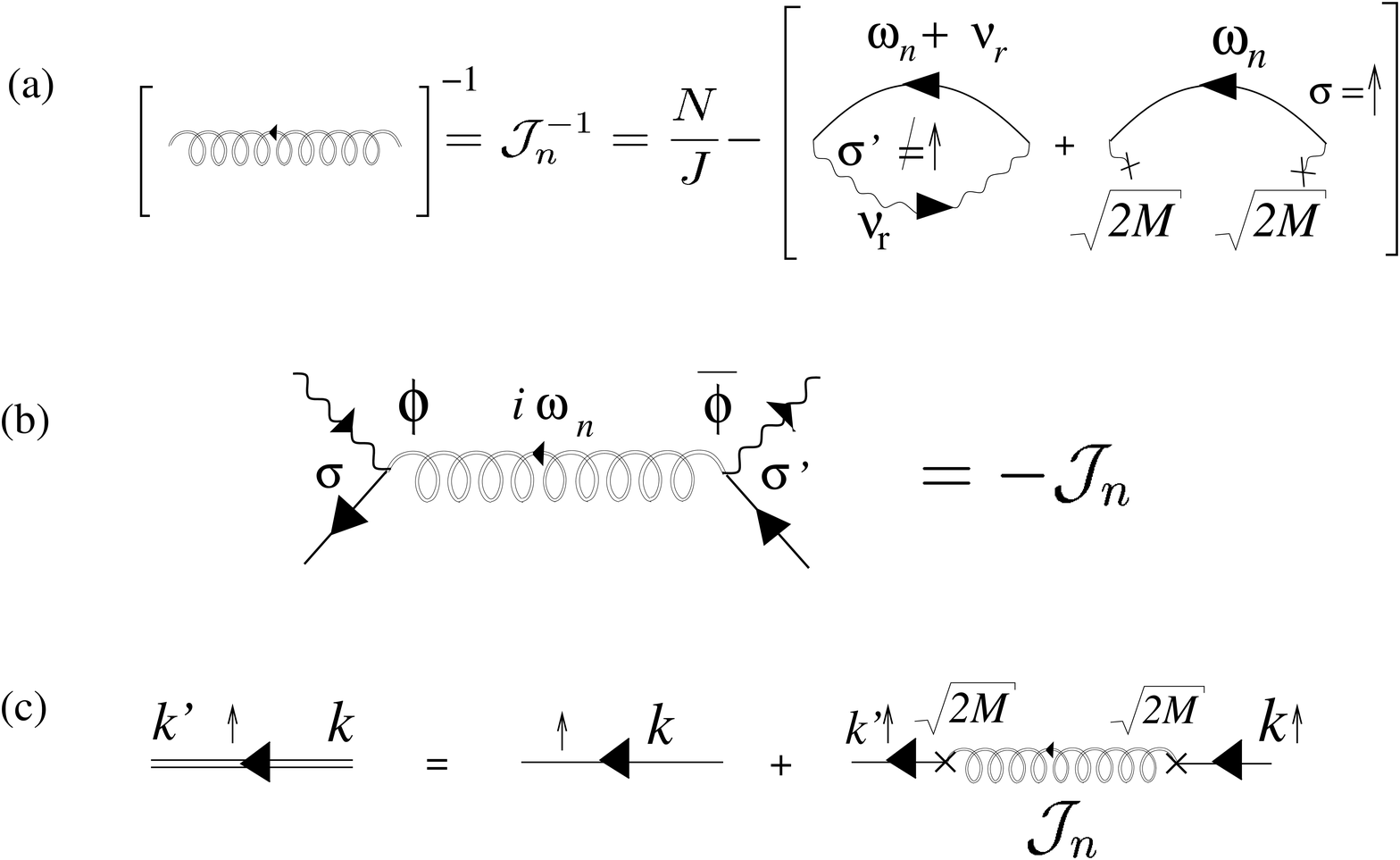}{fig2}{
Feynman diagrams for the large $N$ limit.  
(a)
Feynman diagrams for the $\phi$ propagator,
where full and wavy lines represent conduction electron 
and Schwinger boson propagators  respectively. 
The bubble diagram
involves  a sum over all $\sigma '\neq \uparrow $ and the crosses in
the second diagram denote the 
Schwinger boson condensate $\langle b_{\uparrow}\rangle
=\sqrt{2M}$. 
(b) Effective Kondo interaction is mediated  by the $\phi$ propagator ${\cal J}_n$.
(c) In the polarized phase, the t-matrix of the $\uparrow$ electrons
is determined by the $\phi$ propagator. 
} 
\shift=-0.4cm
When we compute these Feynman diagrams, we find that ${\cal J}_{n}= 
N^{-1}{\cal  J}(i\omega_{n}+\lambda_{0}) $, where
\[
\rho {\cal J}(\omega+i\delta )= \left[
\psi\left(
\frac{1}{2} + \frac{\omega }{2 \pi i T} 
\right)
-\ln \frac{T_K 
}{2\pi i T}  
+ i\pi \tilde{S} \right]^{-1}.
\]
Here $T_{K}=De^{-1/\rho J}$ is the Kondo temperature, expressed in
terms of the band-width $D$ and density of states $\rho $.
This propagator mediates the interaction between the spin bosons 
and electrons, (Fig. \ref{fig2} (b)) 
and describes 
the frequency dependent Kondo coupling constant.
The asymptotic behavior of this function
\[
\rho {\cal  J} (\omega,T) \sim \frac{1}{Log\left(\frac{\max (\omega,2\pi T)}{T_{K}} \right)}
\]
describes a coupling constant which is small and antiferromagnetic
(positive) at high energies, while small  and ferromagnetic (negative)
at low energies. The cross-over from antiferromagnetic behavior at
high energies, to ferromagnetic behavior at low energies, is a
well-known feature of this model\cite{varma,noz,tsvelik,natan,Cragg}. 

By carrying  out the 
Gaussian integral over the fluctuations of the $\phi $ field, 
we are able to compute the correction to the free energy $F_{LM}$ of the free moment
due to the Kondo effect, 
\begin{eqnarray}\label{result1}
F_i=F_{LM} (T,B) + \int 
\frac{d\omega}{\pi}f(\omega)
\alpha \bigl[\omega+ \max (T\zeta,\tilde{B})\bigr]. 
\end{eqnarray}
where   the phase shift
$\alpha (\omega) =  {\rm Im} \
ln\left[ {\cal  J}^{-1}(\omega+i\delta )
\right]$. 
In zero field  $F_{LM}= - TS_{o}$, where 
\begin{eqnarray}\label{}
S_0[\tilde{S}] &=& N [(1+ \tilde{S})\ln (1 + \tilde{S}) -
\tilde{S} \ln \tilde{S} ]\cr&-& \frac{1}{2}\ln [ 2\pi  N
\tilde{S}(1+\tilde{S})]+O (1/N), 
\end{eqnarray} 
is the entropy of a free $SU (N)$ spin $S$. 
In the polarized phase, 
\begin{eqnarray}\label{}
F_{LM}&=& 
(N-1)T\ln (1- e^{-\beta \tilde{B}}) - SB .
\end{eqnarray}
We now use these results to characterize the nature of the excitation
spectrum as a function of field. 

In zero field, the second term in (\ref{result1}) can be
expanded at low temperatures 
as a power-series in the small parameter $g= \rho {\cal J} = 1/Log(2
\pi T/T_{K})$. We find that 
the leading order contribution
to the entropy is  given by 
\begin{eqnarray}\label{}
S_{T=0} &=& S_{0}[\tilde{S}]-\zeta -  {2 \pi^{2}\tilde{S}
(\tilde{S}+1)} g (T)^{3}+O (g^{4})\cr
&=&S_0[\tilde{S} -{\textstyle \frac{1}{N}}]
+ \frac{2 \pi^{2}\tilde{S} (\tilde{S}+1)}
{Log^{3} (\frac{T_{K}}{2\pi
T})} +O (g^{4}).
\end{eqnarray}
From this result, we see that the entropy at low temperatures
is that of a spin $S^* = \frac{N}{2}(\tilde{S} -\frac{1}{N})=
S- \frac{1}{2}$, 
quenched by
one half unit. 
The $g^{3}$ term is the leading 
perturbative correction to the entropy of a Kondo problem, but with  the
bare antiferromagnetic coupling constant $\rho J$ replaced
by the running (ferromagnetic) coupling constant $\rho {\cal J}$. 

If we differentiate $\Delta S$ with respect to temperature, we see
that the logarithm in  $\rho {\cal  J}$
leads to a divergent low-temperature specific heat coefficient (fig \ref{fig3})
\[
\frac{C_{V}}{T} = \frac{\partial S}{\partial T}=
6\pi^{2}\frac{\tilde{S} (\tilde{S}+1)}{T \ln ^{4}\frac{T_{K}}{2\pi T }},
\]
The singular nature of this 
specific heat coefficient 
indicates that the 
fluid of excitations in zero field
is not a Landau Fermi liquid. Moreover, 
we can see that this result is 
actually a generic consequence 
of the singular temperature dependence of the coupling constant.
\figwidth=\columnwidth \fg{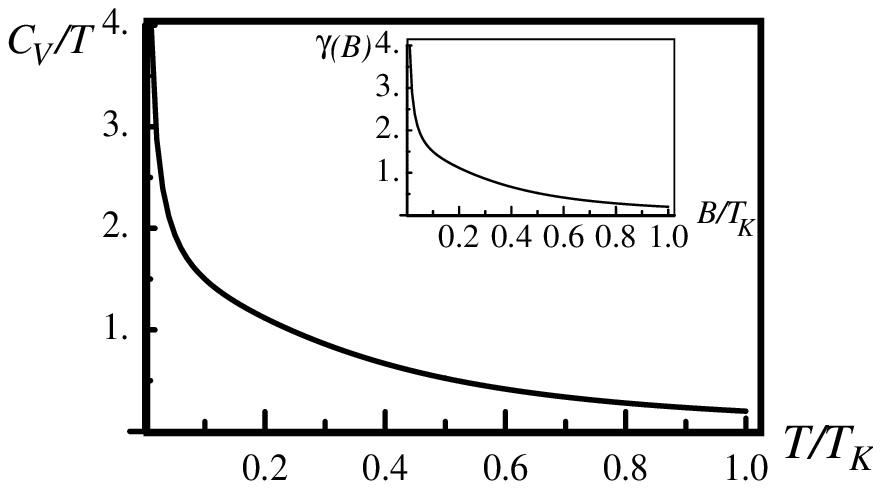}{fig3}{The  zero field
specific heat capacity of the under-screened Kondo model for the case
$\tilde{S} =1/2$, showing the $1/T \ln ^{4} (T_{K}/T)$ divergence
at low temperatures. Inset, the field dependent linear specific heat 
}

To gain further insight into this non-Fermi liquid
behavior, let us examine the effects of a magnetic field. 
When we differentiate the Free energy (\ref{result1})
to obtain the magnetization, 
the combination of frequency and magnetic field 
in the phase shift $\alpha (\omega
+g_{N}B)$, enables us to replace the field derivative with 
 with a frequency derivative inside the integral,  so that at $T=0$,
\begin{eqnarray}\label{}
M &=& - \frac{\partial F}{\partial B} 
= S - 
\frac{g_N}{\pi}
\alpha (\tilde{B})\cr
&=&S-\frac{1}{2\pi}\left(
\frac{\pi}{2}-
\tan^{-1} \left(\frac{ \ln
\left[
\frac{B }{{2T_K}}\right]}
{\pi\tilde{S} }\right)
 \right)
\end{eqnarray}
where we have replaced
$g_N\rightarrow \frac{1}{2}$  in the large $N$ limit.
$M (B)$ evolves from 
$S$ at high fields, to $S- \frac{1}{2}$ at low
fields (Fig. \ref{fig4}), with a weak ferromagnetic correction
$\Delta M \sim {\tilde{S}}/{\ln (T_{K}/B)}$, at low fields.
\shift=-0.7cm\figwidth=\columnwidth \fg{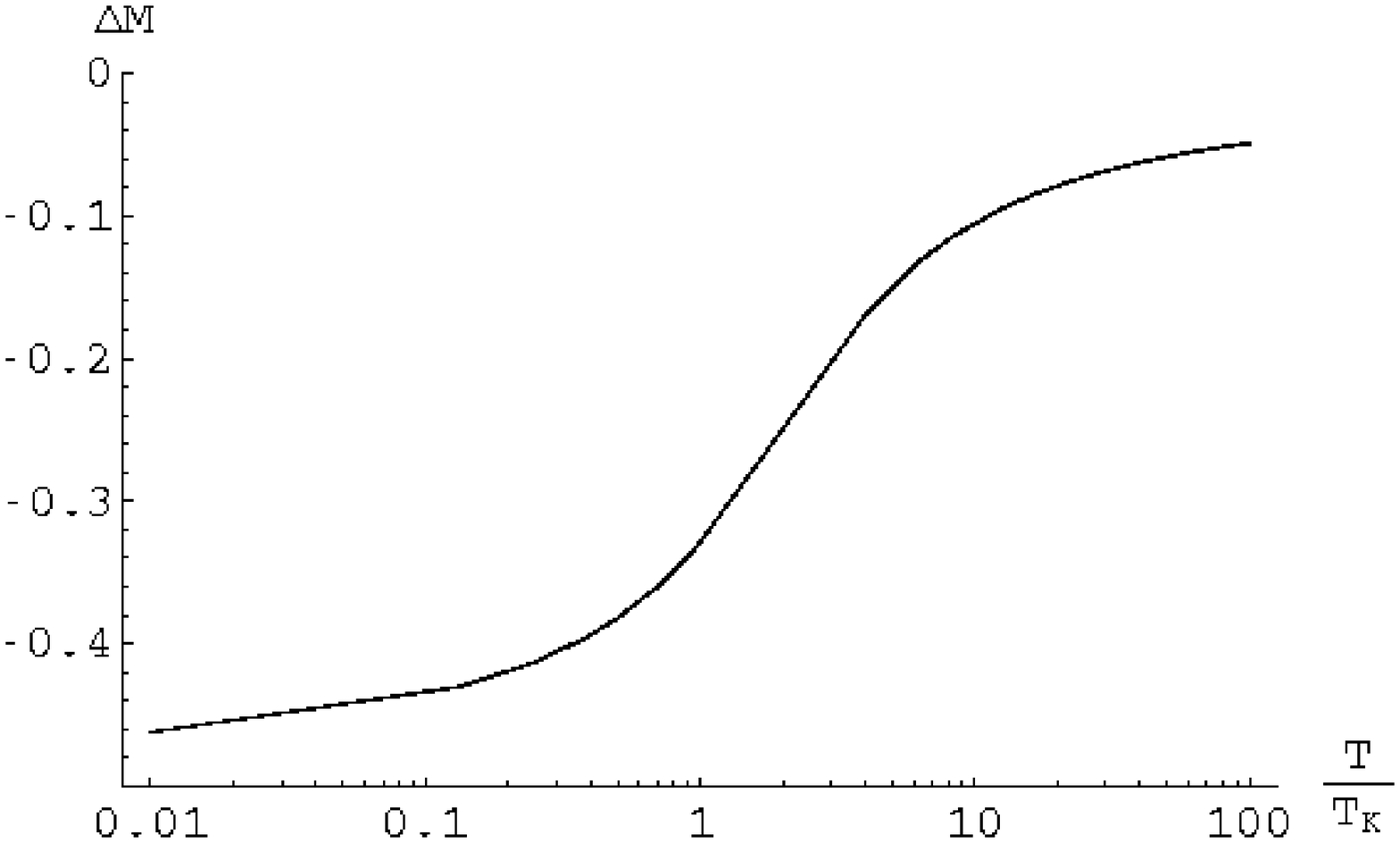}{fig4}{The ground-state
magnetization $\Delta M= S-M$ of the under-screened Kondo model for 
the case $\tilde{S}=0.5$ }

Once the local moment becomes polarized, the specific heat becomes
linear
at low temperatures, and a Fermi liquid is formed. 
In a field at absolute zero, 
the
bose field is condensed with $\langle b_{\uparrow}\rangle =\sqrt{2S}$ so that now
\[
H_{I}\rightarrow 
\sum_{\sigma \neq \uparrow}
\left[
\bar \phi\  b\dg _{\sigma }\psi _{\sigma } +\rm H. c. 
\right]
+
\sqrt{2S}\left(\bar \phi \psi _{\uparrow}+ \rm H. c.  \right)
-\frac{N}{J}\bar \phi \phi , \]
thereby giving rise to a 
{\sl resonant elastic coupling} between the $\phi$ field and the conduction
electrons, so that the t-matrix
$t_{\uparrow} (\omega)$ for the ``up'' 
electrons is now (Fig. \ref{fig2} (c))
\begin{equation}\label{}
t_{\uparrow} (\omega)= \frac{2S}{N}{\cal J} (\omega+\tilde{B}).
\end{equation}
Now since ${\cal  J} = \vert {\cal  J} \vert
e^{-i\alpha (\omega)}$, we can 
identify $\delta_{\uparrow}= -\alpha (B)$ as the 
elastic scattering phase shift of the ``up'' electron at the Fermi surface. 
Furthermore, by linearizing
around $\omega=0$ at zero temperature, we obtain
\begin{equation}\label{}
t_{\uparrow} (\omega-i\delta )= \frac{Z \tilde{S}}{\omega - \xi}
\end{equation}
where $Z=  1/(\partial_{\omega} {\cal  J}^{-1}(\omega))\vert _{\omega=\tilde{B}}
= \tilde{B}/\rho  $ and 
\begin{equation}\label{}
\xi = - Z {\cal J}^{-1} (\tilde{B}-i\delta )= \tilde{B} \ln \frac{T_{K}}{\tilde{B}}+ i \pi \tilde{S}\tilde{B}.
\end{equation}
In other words, the polarized spin generates a resonant scattering
pole of strength $Z=\tilde{B}/\rho $, with phase
shift $\delta _{\uparrow}= -\alpha (\tilde{B})$ and a width $\Delta =\pi \tilde{S }\tilde{B}$
which defines
the characteristic energy scale of the field-tuned Fermi liquid.
The density of quasiparticle states at the Fermi energy associated with this resonance is
given by
\begin{equation}\label{}
N^{*} (0)= \frac{1}{\pi}\alpha ' (\tilde{B}) = \frac{1}{\pi \tilde{B}}
\frac{\pi \tilde{S}}{(\log (\frac{T_{K}}{\tilde{\tilde{B}}}))^{2}+ (\pi
\tilde{S})^{2}}.
\end{equation}
which 
diverges as $(\tilde{B}\log ^{2}(\frac{T_{K}}{\tilde{B}}))^{-1}$ at low fields,
enabling us to read
off the field-dependent Fermi temperature 
\begin{equation}\label{}
T_{F}^{*} \sim N^{*} (0)^{-1}= B \log ^{2}(\frac{T_{K}}{\tilde{B}})
\end{equation}
From this we see that the characteristic energy scale associated with 
Fermi liquid that forms at finite field, is the field itself. At 
temperatures $T\ltappr \tilde{B}$, the specific heat capacity will
become linear, $C_{V}= \gamma (\tilde{B})T$, where $\gamma = \frac{\pi^{2}k_{B}^{2}}{3}N^{*} (0)$ (Fig. \ref{fig3}).
As the field
is reduced to zero, the temperature window for Fermi liquid behavior narrows
to zero,  ultimately vanishing at zero field. 

Certain aspects of these results 
will change at finite $N$.  One of the most important
changes concerns the values of the phase shifts. In the large $N$ limit,
the asymptotic low field limit of the $\delta_{\uparrow}$ phase shift is $-\pi$. 
At finite $N$, by relating the change in 
magnetization to the phase shift, 
$\Delta M= g_N\frac{\delta_\uparrow}{\pi}=- 1/2$ we deduce that 
the asymptotic low field phase shift for the 
``up'' electrons at finite $N$ is 
$\delta_{\uparrow}  = -\pi\left(1 - \frac{1}{N}\right) $.
Since the sum of all the phase shifts for elastic 
spin scattering must equal zero, this implies that  $\delta_{\uparrow} + (N-1)
\delta_{\downarrow}= 0$, or 
$\delta_{\downarrow}= \pi/N$ ($\sigma'\neq \uparrow$).  We see that
modulo the $\pi$ shift, all the phase shifts become equal in the limit
$B\rightarrow 0$, $\delta _{\sigma } {\hbox{mod}} (\pi)= \pi/N$, 
the same phase shift as found in the 
the fully screened Kondo model\cite{noz}.

The field-tuned Fermi liquid in the under-screened Kondo model
follows from general renormalization group
arguments and is expected to extend to all under-screened Kondo
models. 
At fields that are  low compared
with the Kondo temperature $T_K$, the effective ferromagnetic Kondo coupling
constant 
$g\sim \ln \bigl[{T_{K}}/{\hbox{max} (\tilde{B},2\pi
T)}\bigr]^{-1}$ renormalizes slowly to zero, producing 
a small correction to the magnetization
\begin{equation}\label{}
M(B)  = S- \frac{1}{2} +  g (B) + O(g^2).
\end{equation}
Since the Kondo resonance 
is tied to the chemical potential, the field-tuned Fermi liquid 
will have zero charge susceptibility, and following
Nozi{\`e}res and Blandin\cite{noz}, the ratio
between the differential magnetic
susceptibility $\chi (B)= \partial M/\partial B$ and the linear
specific heat $\gamma (B)$ is fixed
$\frac{\chi (B)}{\gamma(B)}=N/(N-1)$,
so that quite generally, 
\begin{equation}\label{}
\gamma^*(B) \sim \chi (B)\sim \frac{1}{B \ln ^2 (B/T_K)}.
\end{equation}
Giamarchi et al. have noted
\cite{noz2} that similar singular behavior 
can also occur in the ferromagnetic Kondo model. 
There is however, an important distinction, for in 
the underscreened Kondo model, the logarithmic factors
above are of order unity, whereas in the ferromagnetic
Kondo model, 
$T_K\rightarrow D e^{+ \frac{1}{J\rho}}>>D$ is exponentially 
larger than the bandwidth, suppressing 
this phenomenon to a small weak-coupling correction.

Our results do not yet give us a  precise understanding of the nature of
the singular Fermi liquid that forms for temperatures $T> B$. 
Our large $N$ treatment suggests that 
the fermionic resonance associated with the binding of the
spin to conduction electron degrees of freedom breaks up
at energy scales above $T\sim B$, as if the heavy quasiparticle
splits up into a ``spinon'' $b$ and a charged spinless ``holon'' $\phi$ in the
non-Fermi liquid phase.  

In conclusion, we have shown how  the treatment of the under-screened
Kondo model using Schwinger bosons enables us to recover the well-known
properties of this model, in the course of which, our results reveal 
a hitherto un-noticed non-Fermi liquid
state at zero field. The model provides an elementary example of a field-tuned
Fermi liquid with a characteristic scale which grows
linearly with the applied magnetic field. 
Intriguingly, 
the low temperature upturn in the specific heat and the
appearance of $B$ as the only scale in the problem are both features
observed in quantum critical heavy electron systems.\cite{schroeder,
gegenwartnew}, leading us 
to speculate that this model may provide a 
useful starting point for future understanding of these 
systems. 

We would like to thank J. Custers, P. Gegenwart and 
F. Steglich for discussions related to this work.
We are particular indebted to G. Zarand and N. Andrei for pointing out
that the phase shift of the $S=1$ under-screened Kondo model is
$\pi/2$, and to C. M. Varma for drawing our attention to reference \cite{noz2}.  This research is partly supported by the National Science
Foundation grant NSF DMR 9983156 (PC).

%-------------------

\vfill \eject

\newpage

%\noindent{\bf Figure Captions}

%\renewcommand{\labelenumi}{{\bf Fig.} \theenumi .}
%\begin{enumerate}

%\item 
%\label{1}

%\item 
%\label{2}

%\end{enumerate}

%\newpage

%\begin{figure}

%\bxwidth=0.6\textwidth
%\prk{full1.ps}{1}

%\caption{}

%\begin{figure}
%\includegraphics[height=5cm]{fig1.eps}
%\end{figure}

%\newpage
%\bxwidth=0.9\textwidth
%\prm{fancyfig3b.eps}{2}

%\caption{}

%\newpage
%\bxwidth =0.9 \textwidth
%\prm{fig3.eps}{3}

%\caption{}

%\caption{}
%\end{figure}
\end{document}